\documentclass[onecolumn,aps,pr,superscriptaddress,preprintnumbers,nofootinbib,10pt]{revtex4-2}
\usepackage{amsmath,amssymb}
\usepackage[dvipdf,dvips]{graphicx}
\usepackage{color}
\usepackage{hyperref}
\usepackage{url}
\usepackage{slashed}
\usepackage{subfigure}
\usepackage{amsmath}
\usepackage{amsfonts}
\usepackage{float} 
\usepackage{epsfig}
\usepackage{graphics}
\usepackage{euscript}
\usepackage{slashed}
\usepackage{epstopdf}
\usepackage[utf8]{inputenc}
\allowdisplaybreaks
\usepackage{pifont}
\usepackage{dsfont}
\usepackage{MnSymbol}
\usepackage{verbatim}
\usepackage{graphicx}
\usepackage{latexsym}
\usepackage{courier}
\usepackage{physics}
\usepackage{bbm}
\usepackage[usenames,dvipsnames]{xcolor}
\usepackage{mathbbol}
\usepackage{mathrsfs}
\usepackage[normalem]{ulem}
\usepackage{bbold}
\allowdisplaybreaks
\usepackage{bbding}
\usepackage{latexsym}
\usepackage{scalerel}

\hypersetup{
	colorlinks=true,
	citecolor=blue,
	citebordercolor=red,
	linktoc=all,
	linkcolor=blue,
	urlcolor=blue
}

\def\eg{{\it e.g.}}
\def\ie{{\it i.e.}}
\def\Det{{\rm Det}}
\def\det{{\rm det}}
\def\tr{{\rm tr}}
\def\Tr{{\rm Tr}}
\def\d{{\rm d}}
\def\Kw{{\stackrel{\mbox{\tiny$\;\bullet$}}{K}}{}}
\def\Rw{{\stackrel{\mbox{\tiny$\;\bullet$}}{R}}{}}
\def\Aw{{\stackrel{\mbox{\tiny$\bullet$}}{\omega}}{}}
\def\Gammabol{{\stackrel{\mbox{\tiny$\circ$}}{\Gamma}}{}}
\def\nablabol{{\stackrel{\mbox{\tiny$\circ$}}{\nabla}}{}}
\def\Rbol{{\stackrel{\mbox{\tiny$\;\circ$}}{R}}{}}
\def\Abol{{\stackrel{\mbox{\tiny$\circ$}}{\omega}}{}}
\def\Dbol{{\stackrel{\mbox{\tiny$\;\circ$}}{D}}{}}

\def\kir{k_{\textmd{IR}}}

\newcommand{\be}{\begin{equation}}
	\newcommand{\ee}{\end{equation}}
\newcommand{\bea}{\begin{eqnarray}}
	\newcommand{\eea}{\end{eqnarray}}

\def\R{\textmd{R}}
\def\L{\textmd{L}}
\def\B{\textmd{B}}

\def\eff{\textmd{eff}}
\def\vol{v_4}
\def\Nf{N_{\textmd f}}

\def\slashD{D\!\!\!\!/\,}
\def\slashDNew{\EuScript{D}\!\!\!\!/\,}
\def\slashA{A\!\!\!/\,}

\newcommand{\barpsi}{\bar\psi}

\newcommand{\tilV}{{\tilde V}}

\def\Vcal{\mathcal{V}}
\def\Acal{\mathcal{A}}

\newcommand{\pd}{\partial}

\definecolor{darkgreen}{rgb}{0.2,0.6,0}
\definecolor{lightblue}{rgb}{0,0.5,0.8}
\definecolor{lightred}{rgb}{0.8,0.2,0.2}
\definecolor{darkorange}{rgb}{1,0.549,0}
\definecolor{brown}{rgb}{0.609, 0.164, 0.164}

\allowdisplaybreaks

\graphicspath{{./}}
\newbox{\ORCIDicon}
\sbox{\ORCIDicon}{\large\includegraphics[width=0.8em]{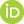}}

\begin{document}
	
	\title{Fate of chiral symmetry in Riemann-Cartan geometry}
	
	\author{Gustavo P. de Brito\,\href{https://orcid.org/0000-0003-2240-528X}{\usebox{\ORCIDicon}}} \email{gustavo@cp3.sdu.dk}
	\affiliation{CP3-Origins, University of Southern Denmark, Campusvej 55, DK-5230 Odense M, Denmark}

	\author{Antonio D. Pereira\,\href{https://orcid.org/0000-0002-6952-2961}{\usebox{\ORCIDicon}}} \email{adpjunior@id.uff.br}
	\affiliation{Institute for Mathematics, Astrophysics and Particle Physics (IMAPP), Radboud University, Heyendaalseweg 135, 6525 AJ Nijmegen, The Netherlands}
	\affiliation{Instituto de F\'isica, Universidade Federal Fluminense, Campus da Praia Vermelha, Av. Litor\^anea s/n, 24210-346, Niter\'oi, RJ, Brazil}
	\author{Arthur F. Vieira\,\href{https://orcid.org/0000-0003-2897-2437}{\usebox{\ORCIDicon}}}\email{afv@cp3.sdu.dk}\email{arthurfv@id.uff.br}
	\affiliation{CP3-Origins, University of Southern Denmark, Campusvej 55, DK-5230 Odense M, Denmark}
	\affiliation{Instituto de F\'isica, Universidade Federal Fluminense, Campus da Praia Vermelha, Av. Litor\^anea s/n, 24210-346, Niter\'oi, RJ, Brazil}

	\begin{abstract}
		We study the mechanism of chiral symmetry breaking for fermionic systems in a gravitational background with curvature and torsion.
		The analysis is based on a scale-dependent effective potential derived from a bosonized version of the Nambu-Jona-Lasino model in a Riemann-Cartan background.
		We have investigated the fate of chiral symmetry in two different regimes. First, to gain some intuition on the combined effect of curvature and torsion, we investigate the regime of weak curvature and torsion. However, this regime does not access the deep infrared limit, which is essential to answer questions related to the mechanism of gravitational catalysis in fermionic systems. Second, we look at the regime of vanishing curvature and homogeneous torsion. In this case, although we cannot probe the combined effects of curvature and torsion, we can access the deep infrared contributions of the background torsion to the mechanism of chiral symmetry breaking. Our main finding is that, in the scenario where only torsion is present, there is no indication of a mechanism of gravitational catalysis. 
	\end{abstract}

	\maketitle
	\section{Introduction}
	\label{Sect:Intro}
	One of the longstanding goals of modern high-energy theoretical physics is the establishment of an ultraviolet-complete theory of quantum gravity which must be internally consistent with the matter degrees of freedom we experimentally observe. In particular, internal consistency demands this would-be quantum gravity theory to be compatible with the observed lightness of Standard Model fermions. 
	
	The simple existence of light fermions (with masses several orders of magnitude smaller than the Planck scale) poses a nontrivial constraint on quantum gravity theories. Since gravity is an attractive interaction, it is conceivable that in a nonperturbative regime, graviton fluctuations could induce the formation of fermionic bound states, possibly inducing chiral symmetry breaking. If such a scenario is realized, one can expect all fermions in nature to have masses of the order of Planck mass, which is incompatible with our Universe. This consistency condition has been tested in the context of asymptotically safe quantum gravity (ASQG)\cite{Niedermaier:2006wt,Codello:2008vh,Reuter:2012id,Percacci:2017fkn,Reuter:2019byg,Eichhorn:2018yfc,Pereira:2019dbn,Reichert:2020mja,Bonanno:2020bil,Pawlowski:2020qer,Dupuis:2020fhh,Eichhorn:2022gku,Saueressig:2023irs} which seems to be compatible with the existence of light fermions~\cite{Eichhorn:2011pc,Meibohm:2016mkp,deBrito:2020dta}. Also, in the context of ASQG, the fate of chiral symmetry was investigated under the impact of topology-changing fluctuations induced by gravitational instantons in a QCD-gravity system \cite{Hamada:2020mug}. For further discussions on the accommodation of fermions in ASQG, see also \cite{Dona:2012am,Dona:2013qba,Eichhorn:2015bna,Meibohm:2015twa,Eichhorn:2016vvy,Biemans:2017zca,Eichhorn:2016esv,Eichhorn:2017eht,Alkofer:2018fxj,Alkofer:2018baq,Eichhorn:2018nda,DeBrito:2019rrh,DeBrito:2019gdd,deBrito:2020xhy,Daas:2020dyo,Daas:2021abx}.
	
	A remarkable aspect of the interplay between gravity and chiral symmetry breaking is the mechanism of \textit{gravitational catalysis}~\cite{Buchbinder:1989fz,Buchbinder:1989ah,Inagaki:1993ya,Elizalde:1993kb,Granda:1995kt,Sachs:1993ss,Elizalde:1995kg,Kanemura:1995sx,Inagaki:1995bk,PhysRevD.56.5097,Geyer:1996wg,Geyer:1996kg,Miele:1996rp,Vitale:1998wm,Inagaki:1997kz,Hashida:1999wb,Gorbar:2007kd,Inagaki:2010py,Sasagawa:2012mn,Gorbar:1999wa,Ebert:2008pc,Hayashi:2008bm,Gorbar:2008sp,Fukaya:2010zz,Flachi:2014jra,Addazi:2017qus,Gies:2018jnv,Gies:2021upb,Gies:2013dca}.
	In this case, chiral symmetry breaking happens thanks to the gravitational background field encoded in spacetime curvature. The phenomenon of gravitational catalysis occurs for fermionic systems in spacetimes with negative curvature~\cite{Buchbinder:1989fz,Buchbinder:1989ah,Inagaki:1993ya,Elizalde:1993kb,Granda:1995kt,Sachs:1993ss,Elizalde:1995kg,Kanemura:1995sx,Inagaki:1995bk,PhysRevD.56.5097,Geyer:1996wg,Geyer:1996kg,Miele:1996rp,Vitale:1998wm,Inagaki:1997kz,Hashida:1999wb,Gorbar:2007kd,Inagaki:2010py,Sasagawa:2012mn,Gorbar:1999wa,Ebert:2008pc,Hayashi:2008bm,Gorbar:2008sp,Fukaya:2010zz,Flachi:2014jra,Addazi:2017qus,Gies:2018jnv,Gies:2021upb,Gies:2013dca}.
	Physically, gravitational catalysis is a consequence of an effective dimensional reduction in the spectrum of the Dirac operator in spacetimes with negative curvature \cite{Gorbar:2008sp}. For example, the low-lying modes of a fermionic system in a $d$-dimensional hyperbolic spacetime behaves effectively as a $(1+1)$-dimensional system like, \eg, in Gross-Neveu or Nambu-Jona-Lasino (NJL) models, which exhibit chiral symmetry breaking~\cite{Gross:1974jv,Nambu:1961tp,Nambu:1961fr}. A renormalization group analysis of the NJL model in an external gravitational field with torsion has been studied in~\cite{shapiro1994interaction}. Gravitational catalysis is analogous to the mechanism of magnetic catalysis occurring in $(2+1)$-dimensional fermionic system in a magnetic background, see, \eg, \cite{Gusynin:1994re,Gusynin:1994va,Scherer:2012nn,Klimenko:2013gua}.

	In \cite{Gies:2018jnv}, the mechanism of gravitational catalysis was used in conjunction with the effects of metric fluctuations in ASQG to establish a bound on the curvature of local patches of spacetime by demanding long-range chiral symmetry to remain intact. As pointed out in \cite{Gies:2018jnv}, the mechanism of gravitational catalysis produces an upper bound on the number of fermions that are compatible with chiral symmetry. This result was later extended to include thermal effects~\cite{Gies:2021upb}.
	
	So far, studies of the phenomenon of gravitational catalysis are mostly restricted to formulations of gravity where the spacetime geometry is fully characterized by the metric. However, there are other proposals of gravitational theories that give rise to richer geometric structures. For example, the general class of metric-affine theories~\cite{Hehl:1994ue,Blagojevic:2013xpa,Baldazzi:2021kaf} treats the spacetime metric and the affine connection as independent objects. In contrast to the metric formulation, where the spacetime geometry can be entirely specified by the curvature tensor, in the metric-affine formulation, the spacetime geometry also depends on information about the nonmetricity and the torsion tensors. The general class of metric-affine theories includes Palatini~\cite{palatini1919deduzione,einstein2005einheitliche,ferraris1982variational} and teleparallel~\cite{Hayashi:1967se,pellegrini1963tetrad,cho1976einstein} theories as special cases.
	
Within metric-affine theories, we can also describe gravity in such a way that the metric and the affine connection are built in terms of more fundamental (and independent) objects, namely, the vierbein and the spin connection. A particular case is the so-called Einstein-Cartan theory, see \cite{Kibble:1961ba,Sciama:1964wt,Hehl:1976kj,Shapiro:2001rz,Trautman:2006fp,Blagojevic:2013xpa}. In general, the spacetime geometry encoded in the Einstein-Cartan formalism is characterized by curvature and torsion. The Einstein-Cartan formulation has the appealing feature of bringing gravity to a language that is closer to the other fundamental interactions. The spin connection is the gauge field associated with local Lorentz symmetry, thus playing a similar role to the gauge bosons in Yang-Mills theories.
	
	The Einstein-Cartan formalism is naturally related to some approaches to quantum gravity, such as loop quantum gravity and spin foams \cite{Perez:2012wv,Rovelli:2014ssa}. In this context, spin foam models formulated in terms of $BF$-theories exhibit a correspondence to the Einstein-Cartan action augmented with the Holst term. The Holst action was also investigated in the context of ASQG, pointing toward the viability of an asymptotically safe theory formulated in terms of Einstein-Cartan formalism \cite{Daum:2010qt}.

	The Einstein-Cartan theory is classically equivalent to the Einstein-Hilbert theory  in the absence of fermions. In the presence of fermions, torsion is naturally sourced and metric and torsion (or more naturally, metric and affine connection) are independent fundamental variables. Indeed, in this case, torsion is, on-shell, expressed in terms of fermionic bilinears. Once it is plugged back into the original action, it gives rise to a metric theory with additional dimension-six operators accounting for the interaction of fermionic axial\footnote{The presence of torsion allows for nonminimal Dirac kinetic terms (see, \eg, \cite{Karananas:2021zkl,Shaposhnikov:2020aen}), which then generate interactions of parity-violating axial-vector currents and vector-vector currents. In this work, we leave out the possibility of considering these nonminimal interactions.} currents~\cite{kibble1961lorentz,Perez:2005pm,Freidel:2005sn}. In this scenario, phenomenological implications for inflation and dark matter production were investigated in~\cite{Shaposhnikov:2020frq,Shaposhnikov:2020aen}. At the quantum level, it is expected that the Einstein-Cartan theory is not equivalent to the quantum Einstein-Hilbert formalism. Thus, in the search for a theory of quantum gravity, it would be helpful to have a classification of the degrees of freedom we should account for in the quantization process.

	In this paper, we explore the possibility of using the interplay between chiral symmetry breaking and gravity to gain some insights on the allowed geometric structures in a gravitational theory due to its coupling to matter degrees of freedom. We focus on the Einstein-Cartan theory, where the spacetime geometry can exhibit nonvanishing torsion. Focusing on a fermionic system coupled to a gravitational background with nonvanishing torsion, we study the impact of the background torsion on the dynamics of four-fermion interactions. In particular, our goal is to understand whether torsion acts in favor or against chiral symmetry breaking, and whether torsion can also act as a gravitational catalyzer. Our findings suggest that spacetime torsion acts in favor of chiral symmetry breaking, but not enough to engender gravitational catalysis.
	
	The paper is organized as follows: In Sec.~\ref{Sect:Setup}, we present the setup of our investigation. In Sec.~\ref{Sect:effective.potential}, we introduce a flow equation for the effective potential, which we use as a tool to investigate torsion effects on chiral symmetry breaking. In Sec.~\ref{Sect:combined.bound}, we present our main result, namely, the impact of a background torsion on the mechanism of chiral symmetry breaking. In Sec.~\ref{Sect:conclusions}, we present our conclusions and outlook. Technical aspects that are relevant to the computations but not essential for the understanding of the main content of this work  are relegated to the Appendix.
	
	\section{Setup \label{Sect:Setup}}
	
	In this section, we present the general setup for the investigation performed in this paper. In this first part, we review the main aspects of the Riemann-Cartan geometry, which is the basis for the Einstein-Cartan theory. In the second part, we introduce the fermionic system used to study chiral symmetry breaking in the presence of curvature and torsion.
	
	\subsection{Overview of Riemann-Cartan geometry  \label{Sect:OverviewRC}}
	In Riemann-Cartan geometry, the basic field variables characterizing the spacetime geometry are the tetrad/vierbein 1-form $e^a(x)={e^a}_{\mu}(x) \,\dd x^\mu$ and the spin connection 1-form $\omega^{ab}(x)=\omega^{ab}_{~~\mu}(x) \,\dd x^\mu$, with $x$ being a generic spacetime point over a manifold $\mathcal{M}$.
	Here and hereafter, the Latin \textit{frame indices}  $a,b,c,\dots$ refer to coordinates in the tangent space $T_x(\mathcal{M})$, while Greek \textit{world indices} $\mu,\nu,\alpha,\dots$ refer to local spacetime coordinates in a given chart. Throughout this paper we work with quantum fields within Euclidean signature\footnote{The use of Euclidean signature is a general limitation of renormalization group methods based on the Wilsonian philosophy that defines a coarse-graining procedure in terms of mass shells, which is an Euclidean concept. See \cite{Manrique:2011jc,Rechenberger:2012dt,Biemans:2016rvp,Saueressig:2023tfy,Horak:2020eng,Bonanno:2021squ,Fehre:2021eob,DAngelo:2022vsh,DAngelo:2023tis} for recent developments on Lorentzian formulation of the Wilsonian renormalization group with a direct or indirect focus toward quantum gravity or quantum fields on curved backgrounds.}, thus $T_x(\mathcal{M})$ correspond to Euclidean flat space. Despite working with Euclidean signature, we will use the term spacetime to refer to the background space where quantum fields are defined. 
	
	For every spacetime point $x \in \mathcal{M}$, a coordinate system $x^\mu$ is related to a local inertial frame $x^a$ in the tangent space by means of the isomorphism $\dd x^a=e^a{}_\mu(x) \dd x^\mu$ and $dx^\mu=e_a{}^\mu(x) dx^a$, with $e_a{}^\mu(x)$ being the inverse of the vierbein field \cite{de1986introduction,bertlmann2000anomalies}. Thus, the metric field arises as a composite field built from the vierbein as $g_{\mu\nu}(x)=e^a{}_\mu(x) e^b{}_\nu(x) \delta_{ab}$, where $\delta_{ab}$ is a flat metric, with Euclidean signature $(+,+,+,+)$.

	The covariant derivative acting on objects with frame (implicit) and world indices is defined by the following rule
	\begin{equation}
		D_\mu A_{\nu_1\dots\, \nu_k}=(\partial_\mu+\omega_\mu)A_{\nu_1\dots\, \nu_k}-\sum_{i=1}^{k}\Gamma^\lambda_{\mu\nu_i}A_{\nu_1\dots\,\nu_{i-1}\lambda \nu_{i+1}\dots\,\nu_k},
	\end{equation}
	where $\omega_\mu=\frac{1}{2}\omega^{ab}_{~~\mu}\Sigma_{ab}$, with $\Sigma_{ab}$ being the generators of the orthogonal group. Both $\omega^{ab}_{~~\mu}$ and $\Sigma_{ab}$ are antisymmetric in the tangent space indices, \textit{i.e.}, $\omega^{ab}{}_\mu=-\omega^{ba}{}_\mu$ and $\Sigma_{ab} =-\Sigma_{ba}$.
	We assume the compatibility conditions on the vierbein and the metric, which are expressed by $D_\mu e^a{}_\nu=0$ and $D_\mu g_{\nu\lambda}=\nabla_\mu g_{\nu\lambda}=0$, where $\nabla_\mu$ is the usual covariant derivative that acts on objects containing only world indices, thus depending only on the affine connection $\Gamma_{\mu\nu}^\alpha$. These relations ultimately ensure that the nonmetricity tensor vanishes and allow us to establish a relation between the affine connection and the spin connection, namely
	\begin{equation}
		\Gamma^\lambda_{\mu\nu}=e_a{}^{\lambda}(\partial_\nu e^a{}_\mu+\omega^a_{~b\nu} e^b{}_\mu).\label{eq:consistencycondition}
	\end{equation}
	
	Equation~\eqref{eq:consistencycondition} implies that the affine connection $\Gamma_{\mu\nu}^\alpha$ is not necessarily symmetric in its lower indices. The antisymmetric part of the connection $\Gamma_{\mu\nu}^\alpha$ defines the torsion tensor ${T^\lambda}_{\mu\nu}={e_a}^{\lambda}{T^a}_{\mu\nu}=\Gamma^\lambda_{\mu\nu}-\Gamma^\lambda_{\nu\mu}$, which can be expressed as the field strength of the vierbein, \ie,
	\begin{align}\label{eq:torsion_def}
		T^a{}_{\mu\nu}=\partial_{\mu}e^a{}_\nu-\partial_{\nu}e^a{}_\mu+\omega^{a}{}_{c\mu}e^c{}_\nu-\omega^{a}{}_{c\nu}e^c{}_\mu.
	\end{align}
	Additionally, the components of the field strength of the spin connection in the dual basis define the Riemann-Cartan curvature given by
	\begin{equation}\label{eq:Riemann-Cartan}
		R^{ab}{}_{\mu\nu}=\partial_{\mu}\omega^{ab}{}_\nu-\partial_{\nu}\omega^{ab}{}_\mu+\omega^{a}{}_{c\mu}\omega^{cb}{}_{\nu}-\omega^{a}{}_{c\nu}\omega^{cb}{}_{\mu}.
	\end{equation}
	Using the vierbein to recast local indices in terms of world indices, we can write $R^{\alpha}_{\,\,\beta\mu\nu} = e^\alpha_{\,\,a} \, e_{\beta \,b}R^{ab}{}_{\mu\nu}$, which is the usual Riemann tensor associated with the affine connection $\Gamma_{\mu\nu}^\alpha$.
	
	Manipulating Eq.~\eqref{eq:torsion_def}, one can solve it for the spin connection, resulting in the following expression
	\begin{equation}\label{eq:spinconnection_decomp}
		\omega^{ab}{}_\mu=\Abol^{ab}{}_\mu(e) +K^{ab}{}_\mu(e,T) .
	\end{equation}
	The first term is the Levi-Civita spin connection, which encodes the torsion-independent part of $\omega^{ab}{}_\mu$. One can fully express the Levi-Civita spin connection in terms of the vierbein, namely
	\begin{equation}
		\Abol^{ab}{}_{\mu}(e)=\frac{1}{2}e_{c\mu}\left(\Omega^{abc}+\Omega^{bca}-\Omega^{cab}  \right),
	\end{equation}
	where $\Omega^{abc}=e^{a\nu}e^{b\lambda}\left(\partial_\nu e^c{}_{\lambda}-\partial_\lambda e^{c}{}_{\nu}  \right)$.
	Throughout this paper, we use the circle on top of geometrical objects to indicate that they are associated with the torsion-independent part of the underlying geometry.
	The second term in \eqref{eq:spinconnection_decomp} is the contorsion tensor, defined by
	\begin{equation}\label{eq:contorsion.torsion}
		K^{ab}{}_\nu=\frac{1}{2}\Big(e^{a\lambda}e^{b\mu}-e^{b\lambda}e^{a\mu}\Big)(T_{\lambda\mu\nu}-T_{\mu\nu\lambda}+T_{\nu\lambda\mu}).
	\end{equation}
	Furthermore, using Eqs. \eqref{eq:Riemann-Cartan} and \eqref{eq:spinconnection_decomp}, we can write the relation
	\begin{align}
		R^\alpha{}_{\beta\mu\nu}=\Rbol^\alpha{}_{\beta\mu\nu}+\nablabol_{\mu}K^\alpha{}_{\beta\nu}-\nablabol_{\nu}K^\alpha{}_{\beta\mu}+K^\alpha{}_{\lambda\mu}K^{\lambda}{}_{\beta\nu}-K^{\alpha}{}_{\lambda\nu}K^{\lambda}{}_{\beta\mu},
	\end{align}
	where ${\mathring{\nabla}}_\mu$ and $\Rbol\,\!^{\lambda}{}_{\rho\mu\nu}$ denote, respectively, the covariant derivative and Riemann curvature associated with the (torsionless) Levi-Civita connection $\Gammabol_{\mu\nu}^\alpha$.
	\subsection{Four-fermion interactions in a Riemann-Cartan framework \label{Sect:Framework}}
	To study the impact of torsion on the mechanism of chiral symmetry breaking, we consider the NJL \cite{Nambu:1961tp,Klevansky:1992qe} for a system with Dirac fermions in a curved  background with (nonvanishing) torsion. We start from a chirally symmetric (Euclidean) action containing the $(\Vcal+\Acal)$-channel of four-fermion interactions
	\begin{equation}\label{eq:full.action1}
		S[\psi,\bar{\psi};g]=\int_x \bigg(\frac{i}{2} 
		\big( \barpsi^i \gamma^\mu D_\mu \psi^i - D_{\mu} \barpsi^i \gamma^\mu \psi^i \big) 
		-\frac{\lambda}{4}(\Vcal+\Acal)\bigg),	
	\end{equation}
	with local four-fermion operators of the form
	\begin{align}
		\Vcal&=(\barpsi^i\gamma_\mu \psi^i)(\barpsi^j\gamma^\mu \psi^j),\\
		\Acal&=(\barpsi^i i\gamma_\mu \gamma_5 \psi^i)(\barpsi^j i\gamma^\mu \gamma_5 \psi^j),
	\end{align}
	where $i,j \in \left\{1,\ldots,\Nf \right\}$, where $\Nf$ is the number of Dirac fermions and the sum over the indices $i,j,\cdots\,$ is implied\footnote{The reader should be careful with the use of Latin indices for both tangent space indices and internal labels for fermionic fields. We use letters from the beginning of the alphabet to denote tangent space indices while for internal labels of Dirac fermions, we take letters from the middle of the alphabet.}.
	Herein, we use $\int_x=\int \dd^4 x \,|e|$ as a shorthand notation for the  integral over the four-dimensional spacetime with $|e|=\det(e^a{}_\mu)$. The Dirac gamma matrices in a curved background are related to their tangent/flat space counterparts through the vierbein, \ie, $\gamma_\mu=e^a{}_\mu \gamma_a$, and satisfy the Clifford algebra $\{\gamma_\mu,\gamma_\nu\}=2 g_{\mu\nu}\mathbf{1}_4$ and $\gamma_5 = \gamma^0 \gamma^1\gamma^2\gamma^3$. 
	The covariant derivative acts on Dirac fermions as follows
	\begin{align}\label{eq:covD_fermion}
		D_\mu \psi & = \partial_{\mu} \psi + \frac{1}{8} \, \omega^{ab}{}_\mu [\gamma_a , \gamma_b ] \, \psi \,, \\
		D_\mu \bar{\psi} & = \partial_{\mu} \bar{\psi} - \frac{1}{8} \omega^{ab}{}_\mu\, \bar{\psi}\, [\gamma_a , \gamma_b ] \,.
	\end{align}
	Self-consistency requires that Euclidean dual spinors are constructed with an extra imaginary factor, namely $\barpsi=(i\psi)^{\dagger}\gamma^0$.
	
	The system \eqref{eq:full.action1} is symmetric under the global chiral group $\textmd{U}(\Nf)_\L\bigotimes \textmd{U}(\Nf)_\R$, corresponding to transformations of the form
	\begin{align}
		&\psi_i \mapsto \psi'_i = (U_\text{R})_{ij} \, P_\text{R} \psi_j + (U_\text{L})_{ij} \, P_\text{L} \psi_j \,,\\
		&\bar{\psi}_i \mapsto \bar{\psi}'_i = \bar{\psi}_{j} P_\text{L} \, (U_\text{R}^\dagger)_{ji}  
		+ \bar{\psi}_{j} P_\text{R} \, (U_\text{L}^\dagger)_{ji} \,,
	\end{align}
	where $U_\text{R,L}^\dagger U_\text{R,L} = U_\text{R,L} U_\text{R,L}^\dagger = 1$ with $P_{\R,\L}=\frac{1}{2}(\mathbf{1}\pm \gamma_5)$.
	Such symmetry can be spontaneously broken if there is a finite condensate formation $\expval{\barpsi \psi}$.
	Chiral symmetry also allows a four-fermion operator of the form $(\Vcal - \Acal)$. However, we restrict our analysis to the $(\Vcal + \Acal)$-channel, which is the channel associated with chiral symmetry breaking. The $(\Vcal - \Acal)$-channel might give rise to a vector condensate \cite{Braun:2011pp} and is left out in the present analysis.
	
	Using Fierz rearrangements, one can recast the $(\Vcal+\Acal)$-channel in terms of scalar and pseudoscalar channels, as follows
	\begin{equation}
		\label{eq:V+A:1}
		\Vcal+\Acal=-2\left[(\barpsi^i \psi^j)(\barpsi^j \psi^i)-(\barpsi^i\gamma_5\psi^j)(\barpsi^j\gamma_5\psi^i)\right].
	\end{equation}
	Decomposing $\psi^i$ in terms of its chiral components, $\psi^i = \psi^i_\R  + \psi^i_\L$  (with  $\psi^i_{\R,\L}=P_{\R,\L}\psi^i$), we can recast \eqref{eq:V+A:1} as
	\begin{equation}
		\Vcal+\Acal=-8(\barpsi^i_\L\psi^j_\R)(\barpsi^j_\R\psi^i_\L).
	\end{equation}
	Thus, the action \eqref{eq:full.action1} turns into
	\begin{equation}
		\label{eq:full.action2}
		S[\psi,\bar{\psi};g]=
		\int_x \bigg(\frac{i}{2} \big( \barpsi^i \gamma^\mu D_\mu\psi^i -D_{\mu}\barpsi^i \,\gamma^\mu \psi^i \big)
		+2\lambda\,(\barpsi^i_\L\psi^j_\R) (\barpsi^j_\R \psi^i_\L)\bigg) \,.
	\end{equation}
	
	To investigate chiral symmetry breaking, it is convenient to consider a partially bosonized version of the four-fermion model in \eqref{eq:full.action2}, which can be obtained by Hubbard-Stratonovich transformation \cite{Braun:2011pp,Inagaki:1997kz}.
	The bosonized action corresponding to \eqref{eq:full.action2} can be written as
	\begin{equation}
		\label{eq:bosonized.action}
		S_\B=\int_x \bigg(\frac{i}{2} \big( \barpsi^i \gamma^\mu D_\mu\psi^i -D_{\mu}\barpsi^i \,\gamma^\mu \psi^i \big)
		+\frac{1}{2\lambda}\tr(\phi^\dagger\phi)+i\bar{\psi}^i\left[P_\L(\phi^\dagger)_{ij}+P_\R \phi_{ij}\right]\psi^j\bigg) \, ,
	\end{equation}
	where we have introduced a conjugate pair of matrix-valued fields $\phi$ and $\phi^\dagger$, which are scalars under Lorentz transformations. 
	We can recover the action \eqref{eq:full.action2} by integrating out $\phi$ and $\phi^\dagger$.
	The Yukawa-like interaction now forces $\phi$ to transform according to
		\begin{align}
			&\phi \mapsto \phi' = U_\text{R} \, \phi\, U_\text{L}^\dagger \,,\\
			&\phi^\dagger \mapsto {\phi'}^\dagger = U_\text{L} \,\phi^\dagger \,U_\text{R}^\dagger \,,
		\end{align}
	so that the partially bosonized action continues to be chirally symmetric ~\cite{Braun:2011pp}. 
	
	In this scenario, the spontaneous breaking of chiral symmetry translates into a finite and positive expectation value $\expval{\phi}$, leading to a mass term for the fermion.
	Thus, the expectation value of the field $\phi$ can be seen as an order parameter. To determine whether or not $\phi$ has a nonvanishing expectation value, we analyze the structure of minima of the effective potential $V_\textmd{eff}(\phi,\phi^\dagger)$ obtained by integrating out fermionic fluctuations.
	
	Since we are interested in the effects of the background torsion on the mechanism of chiral symmetry breaking, it is useful to rewrite \eqref{eq:bosonized.action} in such a way that we can make the torsion contribution explicit.
	Using Eqs. \eqref{eq:spinconnection_decomp} and \eqref{eq:covD_fermion}, one can write
	\begin{align}
		D_\mu \psi&=\Dbol_\mu\psi+K_\mu\psi, \\
		D_\mu \bar{\psi}&=\Dbol_\mu\bar{\psi}-\bar{\psi}K_\mu,
	\end{align}
	with $\Dbol_\mu$ being the Dirac covariant derivative built with the torsionless spin connection and $K_{\mu}=\frac{1}{8}K^{ab}{}_\mu [\gamma_a,\gamma_b]$ is the $\textmd{so}(4)$-valued contorsion, which, using the relation \eqref{eq:contorsion.torsion}, reads
	\begin{equation}\label{eq:contorsion.torsion.2}
		K_{\mu}=\frac{1}{16}\big[\gamma^\alpha,\gamma^\beta\big](T_{\alpha\beta\mu}-T_{\beta\alpha\mu}-T_{\mu\alpha\beta}) \,.
	\end{equation}
	Upon integration by parts, the Dirac term can be expressed as
	\begin{align}\label{eq:Dirac.action1}
		S_{\textrm{Dirac}} = \int_x \frac{i}{2} \big( \barpsi^i \gamma^\mu D_\mu\psi^i -D_{\mu}\barpsi^i \,\gamma^\mu \psi^i \big) = \int_x \left( i\bar{\psi}^i \mathring{\slashD}\psi^i+\frac{i}{2}\bar{\psi}^i\big\{\gamma^\mu,K_{\mu}\big\}\psi^i \right)\,.
	\end{align}
	At this point, it is convenient to decompose the torsion tensor in terms of its irreducible components $T_\mu$ (vector component), $A_{\mu}$ (axial-vector component) and $q_{\mu\nu\rho}$ (irreducible rank-3 component) \cite{Harst:2014vca,Shapiro:2014kma,Karananas:2021zkl}. Explicitly,
	\begin{equation}\label{eq:torsion.decomp}
		T^\lambda{}_{\mu\nu}=\frac{1}{3}(\delta^\lambda_\nu T_\mu-\delta^\lambda_\mu T_\nu)+\frac{1}{6}{\epsilon^\lambda}_{\mu\nu\sigma}A^\sigma + {q^\lambda}_{\mu\nu},
	\end{equation}
	with ${q^\lambda}_{\mu\lambda}=0$, $\epsilon^{\mu\nu\rho\sigma}q_{\nu\rho\sigma}=0$ and ${q^\lambda}_{\mu\nu}=-{q^\lambda}_{\nu\mu}$. Furthermore, $T_\mu={T^\lambda}_{\mu\lambda}$ is the trace of the torsion tensor and $A^\rho={\epsilon_\lambda}^{\mu\nu\rho}{T^\lambda}_{\mu\nu}$ is the axial-trace vector. Using Eqs. \eqref{eq:contorsion.torsion.2} and \eqref{eq:torsion.decomp}, together with the identity $\big\{\gamma^\mu,\big[\gamma^\alpha,\gamma^\beta\big]\big\}=4\epsilon^{\mu\alpha\beta\rho}\gamma_5\gamma_\rho$, yields
	\begin{equation}
		\big\{\gamma^\mu,K_{\mu}\big\}=\frac{1}{4}\gamma_5 \slashA.
	\end{equation}
	Thus, the Dirac action finally turns into
	\begin{equation}
		S_{\textrm{Dirac}} = \int_x \left( i\bar{\psi}^i \mathring{\slashD}\psi^i + \frac{i}{8}\bar{\psi}^i \gamma_5\slashed{A}\psi^i \right) =\int_x i\bar{\psi}^i \slashDNew\psi^i \,,
	\end{equation}
	where we have introduced a new derivative operator defined by
	\begin{align}
		&\EuScript{D}_\mu \psi=\Dbol_\mu\psi-\frac{1}{8}\gamma_5A_\mu \psi \,,\\
		&\EuScript{D}_\mu \bar{\psi}=\Dbol_\mu \bar{\psi}+\frac{1}{8} \bar{\psi} \gamma_5 A_\mu  \,.
	\end{align}
	Thus, the minimal coupling of Dirac fermions with gravity in the presence of nonvanishing torsion is equivalent to fermions minimally coupled to a torsionless curved background, plus an axial interaction through $A_\mu$~\cite{buchbinder1992effective,Obukhov:2014fta}. 
	The torsion-dependent term $\bar{\psi}^i \gamma_5\slashed{A}\psi^i$ has certain similarities with parity-violating terms investigated in the context of theories with Lorentz- and CPT-symmetry violations \cite{Mariz:2007gf,Gomes:2008an,Assuncao:2018jkq}. 
	
	Finally, assuming an homogeneous breaking pattern $\phi_{\ij}=\phi_0\delta_{ij}$, with $\phi_0$ being a constant, the bosonized action \eqref{eq:bosonized.action} can be expressed as
	\begin{equation}
		\label{eq:final.bosonized.action}
		S_\B[\psi,\bar{\psi};\phi_0]=\int_x \left[i\bar{\psi}^i\slashDNew \psi^i+i\phi_0\bar{\psi}^i\psi^i+\frac{\Nf}{2\lambda}\phi_0^2\right].
	\end{equation}
	We are now ready to analyze the effective potential associated with \eqref{eq:final.bosonized.action}.
	\section{Effective potential and its flow equation}\label{Sect:effective.potential}
	The fermions appear as bilinears in the bosonized action \eqref{eq:final.bosonized.action} and, once inserted in the Boltzmann weight of the generating functional, they can be readily integrated out. This ultimately provides an expression for a purely bosonic quantum effective action. 
	The associated effective potential $\tilV_{\eff}(\phi_0)$ is given by
	\begin{equation}\label{eq:effective.potential}
		\begin{aligned}
			\tilV_{\eff}(\phi_0)
			&=\frac{\Nf}{2\lambda}\phi_0^2-\frac{\Nf}{\vol}\log\big[ \Det\big(\slashDNew+\phi_0\big) \big]  \\
			&=\frac{\Nf}{2\lambda}\phi_0^2-\frac{\Nf}{2\vol}\Tr\big[\!\log (-\slashDNew^2+\phi_0^2)\big].
		\end{aligned}
	\end{equation}
	The factor $\vol$ stands for the 4-dimensional spacetime volume\footnote{In the case of a noncompact manifold, an appropriate regularization must be employed.}.
	
	Our goal is to explore how local patches of the spacetime geometry influence the mechanism of chiral symmetry breaking. Therefore, instead of computing the effective potential taking into account all modes of the fermionic field, we perform a coarse-grained analysis in terms of a scale-dependent effective potential. The idea is to introduce a momentum scale $k$ that acts as an infrared regulator, such that the effective potential associated with a scale $k$ (denoted as $\tilV_k(\phi_0)$) only includes effects of fermionic modes with ``momentum''%
	\footnote{More precisely, we define the coarse-graining procedure in terms of the differential operator $-\slashDNew^2$.
		Thus, the scale-dependent potential $\tilV_k(\phi_0)$ includes fermionic fluctuations associated with eigenvalues of $-\slashDNew^2$ that are larger than $k^2$.} %
	larger than $k$. In this sense, the scale-dependent effective potential $\tilV_k(\phi_0)$ probes the effects of local patches of geometry with a characteristic length scale of order $\sim 1/k$.
	
	To define the scale-dependent effective potential we follow a strategy inspired by the regularization scheme used in the functional renormalization group (FRG) \cite{Dupuis:2020fhh,Wetterich:1992yh}.%
	\footnote{Alternatively, one can also define a scale-dependent potential using the proper-time regularization scheme as done in \cite{Gies:2018jnv,Gies:2021upb} in the study of gravitational catalysis in Riemannian geometries.} 
	Here, we regularize \eqref{eq:effective.potential} by the replacement
	\begin{equation}
		\log (-\slashDNew^2+\phi_0^2)\mapsto\log (-\slashDNew^2 + R_k(-\slashDNew^2)+\phi_0^2) \,,
	\end{equation}
	where $R_k(-\slashDNew^2)$ is the FRG regulator function.
	The regulator function $R_k(-\slashDNew^2)$ is defined\footnote{Following~\cite{Pagani:2015ema}, we adopt the so-called type-II regularization, in which the argument of the regulator function is the full Dirac operator squared. 
		This choice is motivated by \cite{Dona:2012am}, where the authors argued that the type-II regulator is more appropriate for the treatment of fermions in a curved background. See also \cite{Daas:2020dyo,Daas:2021abx}.} such that it suppresses quantum fermionic fluctuation contributions based on the spectrum of the effective Dirac operator $-\slashDNew^2$, \ie, $R_k(-\slashDNew^2)$ suppresses modes with eigenvalues lower than $k^2$ (see, \eg, \cite{Percacci:2017fkn,Reuter:2019byg} for general properties of the FRG regulator).
	In general, in the limit $k \to 0$ the regulator $R_k(-\slashDNew^2)$ should vanish, implying that $\tilV_{k=0}(\phi_0) = \tilV_{\eff}(\phi_0)$.
	Throughout this work we use the notation $R_k(-\slashDNew^2)=k^2 r(y)$ (with $y = -\slashDNew^2/k^2$), and we explore the Litim \cite{Litim:2000ci} and exponential shape functions, respectively defined by
	\begin{align}
		&\text{Litim:} \hspace*{-1cm} &r(y) =&\, (1-y) \theta(1-y) \,,\label{eq:shapeLitim} \\
		&\text{Exponential:} \hspace*{-1cm}  &r(y) =&\, \frac{y}{e^y - 1} \,.\label{eq:shapeExp}
	\end{align} 
	
	Based on this FRG regularization scheme, we define the scale-dependent effective potential as 
	\begin{align}
		\tilV_k(\phi_0)=\frac{\Nf}{2\lambda}\phi_0^2-\frac{\Nf}{2\vol}\Tr\big[\log (-\slashDNew^2 + R_k(-\slashDNew^2)+\phi_0^2)\big] \,.
	\end{align}
	We define the flow of $\tilV_k(\phi_0)$ by acting on it with a scale-derivative operator $k\pd_k$, such that
	\begin{align}
		\label{eq:flow.Vk}
		k\pd_k \tilV_k(\phi_0)=-\frac{\Nf}{2\vol}\Tr\Big[ \Big( -\slashDNew^2 + R_k(-\slashDNew^2)+\phi_0^2 \Big)^{-1}\, k\pd_k R_k(-\slashDNew^2)  \Big]\,.
	\end{align}
	The right-hand side of \eqref{eq:flow.Vk} is both ultraviolet- and infrared-finite, as long as $R_k(-\slashDNew^2)$ satisfies all the standard properties of a FRG-regulator \cite{Percacci:2017fkn,Reuter:2019byg}. 
	To compute the effective potential at a given scale $\kir$, we integrate the flow equation \eqref{eq:flow.Vk}, resulting in\footnote{From now on, we omit the subscript on $\phi_0$ and we write simply $\phi$ for constant field configurations. }
	\begin{equation}
		\label{eq:Veff.full.1}
		\begin{aligned}
			\tilV_{\kir}(\phi) &= \frac{\Nf}{2\lambda_{\Lambda}}\phi^2 - \int_{\kir}^{\Lambda}\frac{\dd k}{k}k\pd_kV_k(\phi) \\
			&=\frac{\Nf}{2\lambda_{\Lambda}}\phi^2 + \frac{\Nf}{2\vol} \int_{\kir}^{\Lambda}\frac{\dd k}{k} \Tr\Big[ \Big( -\slashDNew^2 + R_k(-\slashDNew^2)+\phi^2 \Big)^{-1}\, k\pd_k R_k(-\slashDNew^2)  \Big]\,, 
		\end{aligned}
	\end{equation}
	where $\Lambda$ is an ultraviolet cutoff scale, and we used the boundary condition $\tilV_{\Lambda}(\phi)=\Nf\, \phi^2/(2\lambda_{\Lambda})$, with $\lambda_{\Lambda}=\lambda$.
	Since the field-independent part of $\tilV_{\kir}(\phi)$ is irrelevant for the analysis of chiral symmetry breaking, it is convenient to define
	\begin{align}
		V_{\kir}(\phi) = \tilV_{\kir}(\phi) - \tilV_{\kir}(0) \,,
	\end{align}
	which automatically removes divergences that are proportional to $\Lambda^4$.
	
	Before discussing the impact of torsion on the mechanism of chiral symmetry breaking, let us briefly review how to identify chiral symmetry breaking from $V_{\kir}(\phi)$. To simplify the discussion, we first consider the case of flat spacetime. In this case, we can compute the trace in \eqref{eq:Veff.full.1} in Fourier space. Computing $V_{\kir}(\phi)$ in a polynomial expansion around $\phi=0$, we find\footnote{In this example, we use the Litim regulator defined in \eqref{eq:shapeLitim}.}
	\begin{equation}
		\begin{aligned}
			V_{\kir}(\phi)
			&= \frac{\Nf}{2\lambda_{\Lambda}}\phi^2  + \frac{\Nf}{8\pi^2} \int_{\kir}^{\Lambda} \frac{\dd k}{k} \left( \frac{k^6}{k^2+\phi^2} - k^4\right)\\
			&= \frac{\Nf}{2\lambda_{\Lambda}}\phi^2 + \frac{\Nf}{16\pi^2} (\kir^2 - \Lambda^2) \phi^2 + \mathcal{O}(\phi^{4}) \\
			&= \frac{\Nf}{2} \left( \frac{1}{\lambda_{\Lambda}} - \frac{1}{\lambda_{\textmd{cr}}} + \frac{1}{8\pi^2} \kir^2 \right) \phi^2 + \mathcal{O}(\phi^{4}) \,,
		\end{aligned}
	\end{equation} 
	where $\lambda_{\textmd{cr}} = 8\pi^2 \Lambda^{-2}$. 
	We are interested in determining if $\phi = 0$ is a local minimum or a local maximum of $V_{\kir}(\phi)$. 
	If $\phi = 0$ is a local maximum, the structure of $V_{\kir}(\phi)$ implies the existence of at least two degenerate minima with nonvanishing $\phi$,%
	\footnote{
		Assuming that the effective potential is bounded from below (which is necessary for stability reasons), it implies that, if $\phi=0$ is a local maximum, the potential $V_{\kir}(\phi)$ has at least two local minima with nonvanishing $\phi$.
	}
	thus implying chiral symmetry breaking. 
	
	Since $V_{\kir}'(0)=0$, $\phi = 0$ is an extremum of $V_{\kir}(\phi)$. Thus, to determine if $\phi = 0$ is a local minimum or a local maximum, we need to investigate the sign of
	\begin{align}
		V_{\kir}''(0)= \Nf \left( \frac{1}{\lambda_{\Lambda}} - \frac{1}{\lambda_{\textmd{cr}}} + \frac{1}{8\pi^2} \kir^2 \right) \,.
	\end{align}
	If $\lambda_{\Lambda} < \lambda_{\textmd{cr}}$, then the sign of $V_{\kir}''(0)$ is positive for all values of $\kir$, implying that $\phi=0$ is a local minimum of $V_{\kir}(\phi)$ for all $\kir$. However, if $\lambda_{\Lambda} > \lambda_{\textmd{cr}}$, then the sign of $V_{\kir}''(0)$ becomes negative for $\kir < k_{\chi\textmd{SB}}$ (where $k_{\chi\textmd{SB}}^2 = 8\pi^2 (\lambda_{\Lambda} -  \lambda_{\textmd{cr}})/(\lambda_{\Lambda} \lambda_{\textmd{cr}})$), implying that $\phi=0$ is a local maximum of $V_{\kir}(\phi)$ for $\kir < k_{\chi\textmd{SB}}$.
	Thus, for $\lambda_{\Lambda} > \lambda_{\textmd{cr}}$, the potential $V_{\kir}(\phi)$ has nontrivial minima for $\kir < k_{\chi\textmd{SB}}$, indicating that quantum fluctuation can trigger chiral symmetry breaking.
	
	From the discussion above, we see that the sign of $V_{\kir}''(0)$ plays a key role in determining whether or not the system exhibits chiral symmetry breaking.
	Using Eq. \eqref{eq:Veff.full.1}, we can write
	\begin{equation}\label{eq:ddV_full}
		V_{\kir}''(0) = \frac{\Nf}{\lambda_{\Lambda}} - \frac{\Nf}{\vol} \int_{\kir}^{\Lambda}\frac{\dd k}{k} \Tr\Big[ \Big( -\slashDNew^2 + R_k(-\slashDNew^2) \Big)^{-2}\, k\pd_k R_k(-\slashDNew^2)  \Big] \,.
	\end{equation}
	In the next section, we use this equation to study the impact of a nontrivial background geometry on $V_{\kir}''(0)$.

	\section{The impact of torsion on the mechanism of chiral symmetry breaking}\label{Sect:combined.bound}	
	
	In this section we investigate the impact of the background torsion on the mechanism of chiral symmetry breaking.
	Our analysis entails the evaluation of the trace on the right-hand side (rhs) of \eqref{eq:ddV_full}. 
	In general, it requires the knowledge of the spectral properties of the nonminimal operator $-\slashDNew^2$ (c.f., Eq. \eqref{eq:nonMinOp}), thus leading to a complicated problem of spectral geometry.
	To simplify the analysis, we focus on two different regimes:
	\begin{itemize}
		\item $|\,\Rbol\,|/\kir^2 \ll 1$ and $A^2/\kir^2\ll 1$: In this regime, we can use early-time heat kernel expansion to evaluate the trace in \eqref{eq:ddV_full}. This approximation allows us to carry the combined effect of background curvature and torsion. However, it is not applicable in the infrared regime.
		\item $\Rbol_{\mu\nu\alpha\beta} \approx 0$ and $A^2$ is approximately homogeneous: In this regime, we can evaluate the trace in \eqref{eq:ddV_full} without employing the early-time heat kernel expansion. Therefore, this approximation allows us to investigate the impact of background torsion in the deep infrared regime. 
	\end{itemize}

	\subsection{Effects of background curvature and torsion}
	
	Now, we investigate the regime where $\Rbol$ and $A^2$ are small in comparison with the cutoff scale $\kir^2$. In this regime, we can evaluate the trace in \eqref{eq:ddV_full} using standard heat kernel methods based on early-time expansion.
	
	For a generic function of the square of the Dirac operator $\EuScript{W}(-\slashDNew^2)$, the heat kernel expansion reads \cite{Percacci:2017fkn,Reuter:2019byg,buchbinder1992effective,avramidi2000heat}
	\begin{align}
		\label{eq:traco.W}
		\Tr \EuScript{W}(-\slashDNew^2)=\frac{1}{(4\pi)^2}\sum_{n=0}^{\infty}\int_x \,\EuScript{Q}_{2-n}[\EuScript{W}]\,\tr\big[b_{2n}(-\slashDNew^2)\big],
	\end{align}
	where $b_{2n}(-\slashDNew^2)$ denotes the nonintegrated heat kernel coefficients for the operator $-\slashDNew^2$. The $\EuScript{Q}$-functionals are defined as 
	\begin{align}
		\EuScript{Q}_n[\EuScript{W}]=\frac{(-1)^p}{\Gamma(n+p)}\int_0^\infty \d z \,z^{n+p-1}\frac{\dd^p\EuScript{W}(z)}{\dd z^p}, 
	\end{align}
	where $p$ denotes some arbitrary positive integer satisfying the restriction $n+p>0$. In particular, if $n$ is positive, then $p=0$. Here, $\tr$ denotes the trace over the internal and spacetime indices. 
	
	For the trace that we are interested in computing (see Eq. \eqref{eq:ddV_full}), we can identify the function
	\begin{equation}
		\EuScript{W}(z) = \frac{k \partial_k R_k(z)}{\big(z + R_k(z)\big)^2} \,,
	\end{equation}
	leading to
	\begin{equation}
		\EuScript{Q}_n[\EuScript{W}] = \frac{1}{\Gamma(n)}\int_0^\infty \d z \,z^{n-1} \frac{k \partial_k R_k(z)}{\big(z + R_k(z)\big)^2} = 2 \,k^{2(n-1)}\mathcal{I}_n[r]\,.
	\end{equation}
	We have introduced the dimensionless threshold integral $\mathcal{I}_n[r]$, which one defines in terms of the shape function $r(y)$ according to
	\begin{equation}
		\mathcal{I}_n[r]=\frac{1}{\Gamma(n)}\int_0^\infty \d y \,y^{n-1} \,\frac{r(y) - y \,r'(y)}{\big(y + r(y)\big)^2} \,.
	\end{equation}
	The numerical value of $\mathcal{I}_n[r]$ depends on the explicit form of the shape function $r(y)$, except in the case $n=1$ where one can show that $\mathcal{I}_1[r] = 1$ for all suitable choices of regulator (see, \eg, \cite{Reuter:2019byg,Percacci:2017fkn}). Since we are interested in the regime where $|\,\Rbol\,|/\kir^2 \ll 1$ and $A^2/\kir^2\ll 1$, we will only keep terms that are at most linear in $\Rbol$ and $A^2$. In such case, we just need to evaluate $\mathcal{I}_2[r]$, which results in $\mathcal{I}_2[r] = 1/2$ for the Litim regulator and $\mathcal{I}_2[r] = 1$ for the exponential regulator.
	
	In order to use the (nonintegrated) heat-kernel coefficients $b_{2n}$ available in the literature, it is useful to rewrite the Dirac operator $-\slashDNew^2$ in a minimal form. In fact,
	\begin{align}\label{eq:nonMinOp}
		-\slashDNew^2=-\Dbol^2 +\hat{B}^\mu\Dbol_\mu+\hat{X} \,,
	\end{align}
	where the operators $\hat{B}^\mu$ and $\hat{X}$ are defined as follows
	\begin{align}
		\hat{B}^\mu&=-\frac{i}{4}\gamma_5\sigma^{\mu\nu}A_\nu \,,\\
		\hat{X}&=\frac{1}{4}\left(\Rbol+\frac{1}{16}A^2\right)\mathbf{1}+\frac{1}{8}(\nablabol\cdot A-i\sigma^{\mu\nu} \nablabol_\mu A_\nu)\gamma_5 \,.
	\end{align}
	The heat-kernel coefficients for this class operators are available, \eg,~in \cite{Obukhov:1983mm,Gusynin:1990ek,Barth:1985cz}. In our analysis, we use
	\begin{eqnarray}
		\tr\big[b_{0}(-\slashDNew^2)\big] &=& 4 \,,\\
		\tr\big[b_{2}(-\slashDNew^2)\big] &=&\frac{2}{3} \Rbol + \frac{1}{2}\tr(\nablabol_\mu \hat{B}^\mu) - \frac{1}{4} \tr(\hat{B}_\mu\hat{B}^\mu) - \tr(\hat{X}) \,.
	\end{eqnarray}
	From the definition of $\hat{B}^\mu$ and $\hat{X}$, one can show that $\tr(\nablabol_\mu \hat{B}^\mu) = 0$, $\tr(\hat{B}_\mu\hat{B}^\mu) = -\frac{3}{4} A^2$, and 
	$\tr(\hat{X}) = \Rbol + \frac{1}{16} A^2$, resulting in
	\begin{align}
		\tr\big[b_{2}(-\slashDNew^2)\big] = -\frac{1}{3} \Rbol + \frac{1}{8} A^2 \,.
	\end{align}
	
	Plugging those results in Eq. \eqref{eq:ddV_full} leads to
	\begin{equation}\label{eq:ddV_Result1}
		V_{\kir}''(0) = \Nf \left(\frac{1}{\lambda_{\Lambda}}  - \frac{1}{\lambda_{\textmd{cr}}} \right)  + \frac{\Nf}{4\pi^2} \mathcal{I}_2[r] \kir^2
		+ \frac{\Nf}{16\pi^2} \left(\frac{1}{3} \langle\Rbol\rangle - \frac{1}{8} \langle A^2 \rangle\right)\,\log(\Lambda^2/\kir^2) \,,
	\end{equation}
	where we have defined spacetime averaged quantities as $\langle (\cdots)\rangle = \frac{1}{v_4} \int_x (\cdots)$.
	Here, we also use the critical coupling $\lambda_\textmd{cr}$ defined as
	\begin{equation}
		\frac{1}{\lambda_\textmd{cr}} = \frac{\Lambda^2}{4\pi^2}\mathcal{I}_2[r].
	\end{equation}
	In order to absorb the logarithmic divergence in the last term of \eqref{eq:ddV_Result1}, we add the following counterterm to the original action
	\begin{equation}
		\delta S[\phi] = \frac{\Nf \,\xi_{\Lambda}}{3}  \int_x \left( R  - T_{\alpha\beta\mu} T^{\alpha\beta\mu} + T_{\alpha\beta\mu} T^{\beta\alpha\mu} + T^{\mu}_{\,\,\,\mu\alpha} T^{\nu \alpha}_{\quad\!\nu} + 2 \,\nabla^{\mu} T^{\nu}_{\,\,\,\,\nu\mu} \right) \,\phi^2 \,,
	\end{equation}
	where all geometrical objects are defined with respect to the full connection $\Gamma_{\mu\nu}^\alpha$. 
	Such a counterterm leads to the following contribution to the scale-dependent effective potential
	\begin{equation}
		\delta V_{\kir}(\phi) = \Nf \,\xi_{\Lambda} \left( \frac{1}{3}  \langle\Rbol\rangle  -  \, \frac{1}{8}   \langle A^2 \rangle  \right) \,\phi^2 \,.
	\end{equation}
	Thus, we define a renormalized effective potential such that
	\begin{equation}
		\label{eq:ddV_Result2}
		V_{\kir}''(0) = \Nf\, \left(\frac{1}{\lambda_{\Lambda}}  - \frac{1}{\lambda_{\textmd{cr}}} \right)  + \frac{\Nf}{4\pi^2} \mathcal{I}_2[r] \kir^2
		+ \Nf \,\xi_{\textmd{IR}} \left(\frac{2}{3} \langle\Rbol\rangle - \frac{1}{4} \langle A^2 \rangle\right)  \,,
	\end{equation}
	where we have introduced the renormalized coupling
	\begin{align}
		\xi_{\textmd{IR}} = \xi_{\Lambda} + \frac{1}{32\pi^2} \,\log(\Lambda^2/\kir^2)  \,.
	\end{align}
	In principle, $\xi_{\textmd{IR}}$ needs to be fixed by experiments. In our analysis, we leave $\xi_{\textmd{IR}}$ as a free parameter.
	
\begin{figure}[!t]
		\begin{center}
			\hspace*{-.5cm}
			\includegraphics[height= 7.cm]{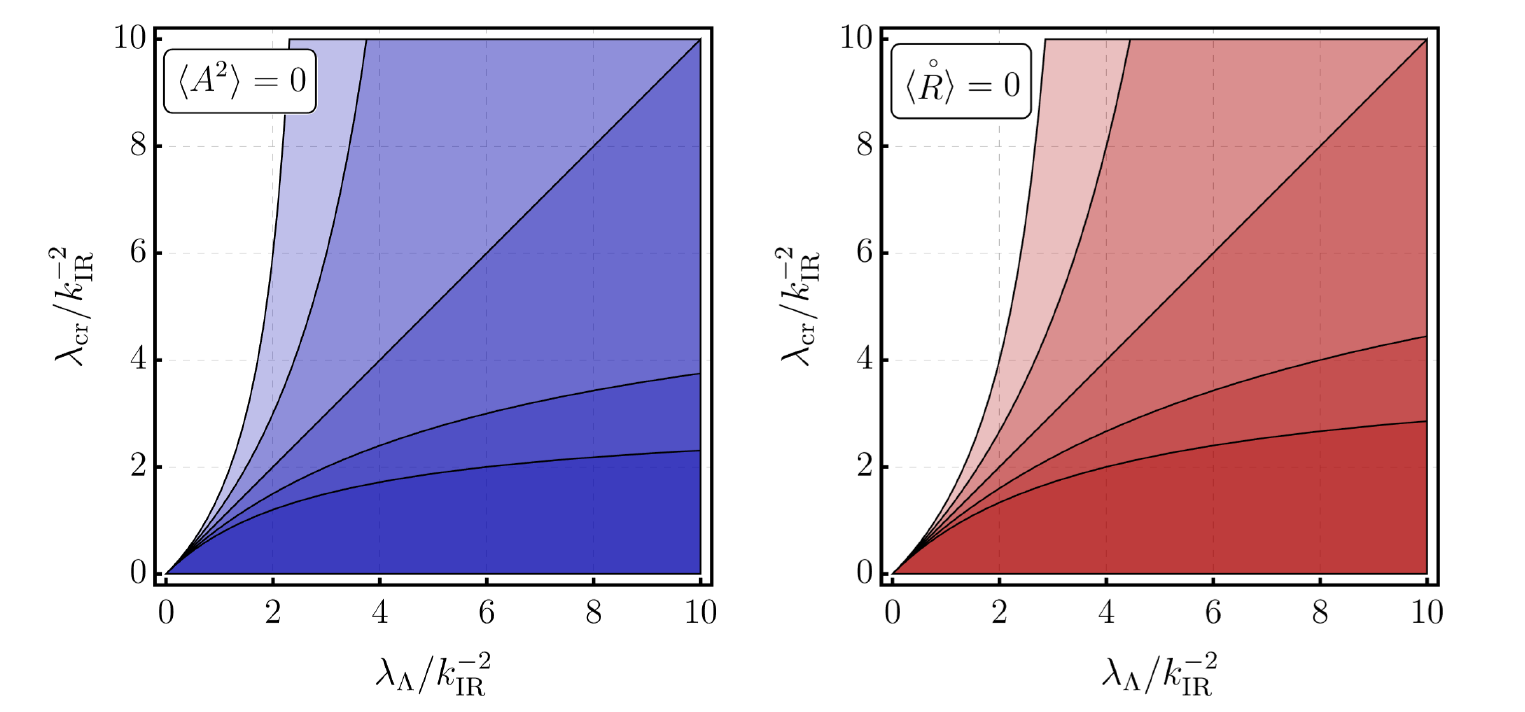}
			\caption{Regions in the parameter space $\lambda_\Lambda \times \lambda_\text{cr}$ where chiral symmetry breaking is triggered by quantum fluctuations in the presence of curvature or torsion. Left panel: we have set $\langle A^2 \rangle = 0$ and we have probed the range $\xi_{\textmd{IR}}\kir^{-2} \langle \mathring{R} \rangle = -1/2\,,-1/4\,,0\,,1/4\,,1/2$ from lighter to darker regions. Right panel: we have set $\langle \mathring{R} \rangle=0$ and from darker to lighter we have probed the range $\xi_{\textmd{IR}}\kir^{-2}\langle A^2 \rangle= -1\,,-1/2\,,0\,,1/2\,,1$.}
			\label{fig::Regions_chSB}
		\end{center}	
	\end{figure}

	As discussed in the previous section, the analysis of chiral symmetry breaking relies on the sign of $\Nf^{-1} V_{\kir}''(0)$. 
	If chiral symmetry breaking is present, there must be a transition scale $k_{\textmd{tr}}$ such that $V_{k_{\textmd{tr}}}''(0)=0$. From \eqref{eq:ddV_Result2}, we find the transition scale
	\begin{equation}
		k_{\textmd{tr}}^2 = \frac{4\pi^2}{\mathcal{I}_2[r]}
		\bigg( \frac{1}{\lambda_{\textmd{cr}}} - \frac{1}{\lambda_{\Lambda}} - \frac{2}{3} \xi_{\textmd{IR}} \langle\Rbol\rangle + \frac{1}{4} \xi_{\textmd{IR}} \langle A^2 \rangle \bigg)\,.
	\end{equation}
	For vanishing curvature and torsion, such transition scale only makes sense if $\lambda_{\Lambda} \geq \lambda_\textmd{cr}$.
	For nonvanishing curvature and/or torsion, the viability of chiral symmetry breaking depends on the inequality
	\begin{equation}\label{eq:Ineq_chSB}
		\frac{1}{\lambda_{\textmd{cr}}} - \frac{1}{\lambda_{\Lambda}} - \frac{2}{3} \xi_{\textmd{IR}} \langle\Rbol\rangle + \frac{1}{4} \xi_{\textmd{IR}} \langle A^2\rangle \geq 0.
	\end{equation} 
	For $\xi_{\textmd{IR}} > 0$, we find that positive (negative) values of $\langle\Rbol\rangle$ act against (in favor of) chiral symmetry breaking, while the axial-torsion term $ \langle A^2\rangle $ acts exclusively in favor of chiral symmetry breaking. For $\xi_{\textmd{IR}} < 0$, the term $\langle\Rbol\rangle$ can still act in favor of or against chiral symmetry breaking (depending on its sign), while the axial-torsion term only acts toward the preservation of chiral symmetry. A comment is in order: due to our use of a Euclidean setting, the sign of $A^2$ is positive definite. In a Lorentzian framework, this is not necessarily true, and a spacetime average can be rather misleading quantity in order to make concrete statements. In such a quantum-field theoretic on a fixed background setting, one can choose specific backgrounds where one can compute quantities beyond their averages.
	In Fig.~\ref{fig::Regions_chSB}, we show the regions in the parameter space $\lambda_\Lambda \times \lambda_\text{cr}$ where chiral symmetry breaking is triggered by quantum fluctuations in the presence of curvature or torsion.
	
	Since the results presented in this section were based on the early-time heat kernel truncated at the first order in $|\,\Rbol\,|/\kir^2$ and $A^2/\kir^2$, our results do not allow us to extract information concerning the deep infrared regime. It is of great importance to determine whether or not there is a mechanism of gravitational catalysis related to a nontrivial background structure. For Riemannian manifolds (\ie, without torsion), the mechanism of gravitational catalysis was investigated, \eg, in \cite{Gies:2018jnv}. If we take such infrared contributions into account, we expect the inequality \eqref{eq:Ineq_chSB} to have an extra term, namely
	\begin{equation}
		\frac{1}{\lambda_{\textmd{cr}}} - \frac{1}{\lambda_{\Lambda}} - \frac{2}{3} \xi_{\textmd{IR}} \langle\Rbol\rangle + \frac{1}{4} \xi_{\textmd{IR}} \langle A^2\rangle 
		+ \mathcal{F}_{\kir} (A_\mu,\Rbol_{\mu\nu\alpha\beta}) \geq 0 \,,
	\end{equation}
	where $\mathcal{F}_{\kir}$ is a function of invariants built from $A_\mu$ and $\Rbol_{\mu\nu\beta\alpha}$ (and their derivatives), and it encodes the contributions beyond the early-time heat kernel expansion.
	We expect that $\mathcal{F}_{\kir}$ acts as the dominant contribution in the infrared. Thus, if $\mathcal{F}_{\kir} > 0$, it would necessarily trigger chiral symmetry breaking. 
	Determining the form of $\mathcal{F}_{\kir}$ associated with the operator $-\slashDNew^2$ (see Eq. \eqref{eq:nonMinOp})  in the presence of curvature and torsion is a complicated problem of spectral geometry lying outside the scope of this paper. In the next section, we show that $\mathcal{F}_{\kir} = 0$ if we neglect curvature effects and restrict the axial-torsion component to be homogeneous.

	\subsection{Is there a torsion-based gravitational catalysis?}
	
	In this section, we study the regime of vanishing Riemannian curvature $\Rbol_{\mu\nu\alpha\beta}=0$ and homogeneous axial-torsion $A_{\mu}$.
	Our goal is to derive the contribution of $A_{\mu}$ to $V''_{\kir}(0)$ beyond the early-time heat kernel approximation, aiming at an understanding if torsion can be a source of gravitational catalysis.
	
	Within the approximation where $\Rbol_{\mu\nu\alpha\beta}=0$ and $A_{\mu}$ is homogeneous, the differential operator $-\slashDNew^2$ reduces to
	\begin{align}\label{eq:DNew_Rzero}
		-\slashDNew^2 = \left( - \partial^2   + \frac{1}{64} A^2 \right) \textbf{1} - \frac{i}{4} \gamma_5 \sigma^{\mu\nu}A_{\nu} \partial_\mu \,.
	\end{align}
	This approximation is useful because it allows us to evaluate the full heat-kernel trace (see Appendix \ref{app:heat_kernel_comput} for more details), resulting in the following expression
	\begin{equation}
		\Tr\left[ \exp(- \tau \,\slashDNew^2) \right] = \frac{1}{16\pi^2 \,\tau^2} \left( 4 + \frac{\tau}{8} A^2 \right) \,.
	\end{equation}
	Surprisingly, the axial-torsion contributes at most up to linear order in $A^2$. This result implies that we can determine all the heat kernel coefficients $b_{n}$ associated with the operator in \eqref{eq:DNew_Rzero}, leading to
	\begin{align}
		\tr [b_{n}(-\slashDNew^2)] = 4 \delta_{n,0} + \frac{1}{8} A^2 \delta_{n,2} \,.
	\end{align} 
	Thus, only $b_{0}$ and $b_{2}$ lead to nonvanishing contributions to the trace in \eqref{eq:traco.W}. In other words, we can truncate the series in \eqref{eq:traco.W} at $n=1$ without resorting to any form of early-time approximation. 
	Based on these results, we find
	\begin{equation}
		\label{eq:ddV_Rzero}
		V_{\kir}''(0) = \Nf\, \left(\frac{1}{\lambda_{\Lambda}}  - \frac{1}{\lambda_{\textmd{cr}}} \right)  + \frac{\Nf}{4\pi^2} \mathcal{I}_2[r] \kir^2
		- \frac{1}{4}  \Nf \,\xi_{\textmd{IR}}  \langle A^2 \rangle  \,,
	\end{equation}
	which correspond exactly to the limit $\Rbol \to 0$ of \eqref{eq:ddV_Result2}. However, the current result is not restricted to small values of $A^2/\kir^2$. 
	
	The most striking feature of Eq.~\eqref{eq:ddV_Rzero} is the absence of torsion-dependent contributions related to the deep infrared regime.
	On physical grounds, a background torsion would affect the mechanism of chiral symmetry breaking by deforming the region in the parameter space of $\lambda_\Lambda \times \lambda_{\kir}$ where we can flip the sign of $V_{\kir}''(0)$. However, our results show no indication of a mechanism of gravitational catalysis based on a background torsion. If $\lambda_\Lambda$ is sufficiently small, then $V_{\kir}''(0)$ remains positive in the deep infrared, thus avoiding chiral symmetry breaking. This is different from the mechanism of curvature-based gravitational catalysis investigated in \cite{Gies:2018jnv}, where chiral symmetry breaking might be triggered even for arbitrarily small values of $\lambda_\Lambda$.
	
	This conclusion may change if we drop the assumptions we considered in this section. It is possible that by combining the effects of curvature and torsion, a more sophisticated geometric-driven mechanism could lead to gravitational catalysis. It is also conceivable that a nonhomogeneous $A_\mu$ could change our conclusion. The analysis in this direction, however, goes beyond the scope of this paper.

	\subsection{Comments on the teleparallel theory}
	
	Manifolds with torsion play an important role in the formulation of teleparallel theories of gravity \cite{Hayashi:1967se,pellegrini1963tetrad,cho1976einstein} (see also \cite{Arcos:2004tzt, Aldrovandi:2013wha,Bahamonde:2021gfp} for reviews).
	This class of theories is a particular case of the general class of metric-affine gravity (MAG)~\cite{Hehl:1994ue,Blagojevic:2013xpa}, and they are known to be classically equivalent to general relativity \cite{Baldazzi:2021kaf}.
	
	The teleparallel formulation is characterized by vanishing curvature and nonmetricity tensors, such that spacetime degrees of freedom are entirely encoded in the torsion tensor.
	In this framework, there is a special spin connection configuration $\Aw^{\,\!ab}{}_{\mu}$, called the Weitzenböck spin connection, that ensures that $\Rw\,\!^{\rho}{}_{\lambda\mu\nu}=0$, with
	\begin{equation}
		\label{eq:omegaw}
		\Aw^{ab}{}_{\mu}=\Abol^{ab}{}_{\mu}+\Kw^{ab}{}_{\mu}.
	\end{equation}
	Here, we use a filled ring to indicate that the geometric quantities are computed with the Weitzenböck spin connection. 
	
	In the following, we briefly discuss the mechanism of chiral symmetry breaking (due to background effects) in the context of teleparallel theories. The key point for such discussion is to identify the prescription of minimal coupling between fermions and gravity in teleparallel theories. Following Refs.~\cite{Krssak:2018ywd,Casadio:2021zai}, in the context of teleparallel theories, fermions couple to gravity according to
	\begin{equation}
		\partial_{\mu}\psi \mapsto \partial_{\mu}\psi+\frac{1}{8}(\Aw^{ab}{}_{\mu}-\Kw^{ab}{}_{\mu})\comm{\gamma_a}{\gamma_b}\psi.
	\end{equation}
	Using condition (\ref{eq:omegaw}) we recover the usual Levi-Civita covariant derivative and no torsion contribution appears. Therefore, the mechanism of chiral symmetry breaking in the context of teleparallel theories is equivalent to the case of Riemannian manifolds explored in \cite{Gies:2018jnv}.

	\section{Conclusions and outlook}\label{Sect:conclusions}
	
	In this work, we have investigated the impact of nontrivial background on the mechanism of chiral symmetry breaking. In particular, we focused on the impact of background torsion.
	Our analysis is based on the evaluation of a scale-dependent effective potential in the bosonized version of the NJL model on a Riemann-Cartan manifold. Within this setting, we used FRG-inspired tools to define a coarse-grained effective potential.
	
	We have analyzed the impact of torsion in two different situations. First, in the approximation where $|\,\Rbol\,|/\kir^2 \ll 1$ and $A^2/\kir^2\ll 1$, we investigated the combined impact of torsion and curvature. 
	In this case, torsion may contribute toward or against chiral symmetry breaking, depending on the infrared value of the renormalized nonminimal coupling $\xi_{\textmd{IR}}$. In fact, for a given sign of the nonminimal coupling $\xi_{\textmd{IR}}$, the torsion contribution plays the analog role of negative curvature, \ie, favors chiral symmetry breaking for positive nonminimal coupling and prevents it for negative coupling.
	The analysis with $|\Rbol \,|/\kir^2$ and $A^2/\kir^2$ does not capture the deep infrared regime. Thus, it does not allow us to investigate a possible torsion-based gravitational catalysis.
	
	The second analysis we performed was in the regime of vanishing curvature and homogeneous torsion. Although this analysis does not capture the combined effects of curvature and torsion, it allows us to investigate the impact of torsion on the mechanism of chiral symmetry breaking in the deep infrared. Surprisingly, in this regime, the only contribution of torsion to $V''_{\kir}(0)$ comes from the leading order correction in $A^2$ in an early-time heat kernel expansion. In physical terms, we have found no indication of a torsion-based gravitational catalysis mechanism.
	
	To our knowledge, this is the first paper investigating the effect of non-Riemannian structures on the mechanism of gravitational catalysis. The results presented here are in agreement with the very recent account~\cite{Vale:2023xlp}, where the possibility of chiral symmetry breaking was also investigated in a background with curvature and torsion, but computing the effective action from the anomaly-induced vacuum effective action in non-Riemannian spacetimes probed in~\cite{Camargo:2022gcw}. Furthermore, the results presented here were restricted to a nondynamical background. As a next step, we aim to investigate the impact of torsion fluctuations on the mechanism of chiral symmetry breaking. In particular, one can investigate the compatibility of light fermions \cite{Eichhorn:2011pc} with quantum gravity scenarios with fluctuating torsion field. We hope to report on this soon.
	
	Torsion effects can also play a role in low-energy physics. For instance, effects of torsion can be emulated in condensed matter systems \cite{kondo1952geometrical,bilby1955continuous,Zubkov:2015cba}. In the context of the geometric theory of defects, the appearance of torsion and curvature in solids are associated with topological defects known as dislocations and disclinations, respectively. Crystalline structures are then viewed as a manifold endowed with a Riemman-Cartan-like geometry.
	The methods used in this paper can, in principle, be adapted to the study of chiral symmetry breaking in low-energy systems that emulate torsion effects. This path could pave the road to the experimental realization of the results presented in this paper by means of analog gravity systems.
	
	
	\bigskip
	\section*{Acknowledgments}
	The authors thank Reinhard Alkofer, Axel Maas, Holger Gies and Aaron Held for fruitful discussions and feedback.
	G.P.B is supported by VILLUM FONDEN	under Grant No. 29405. 
	A.D.P acknowledges CNPq under the grant No. PQ-2 (312211/2022-8), FAPERJ under the “Jovem Cientista do Nosso Estado” program (E26/202.800/2019 and E-26/205.924/2022), and NWO under the VENI Grant (VI.Veni.192.109) for financial support. 
	The work of A.F.V is supported by CNPq under the Grants No. 140968/2020-2 and 200442/2022-8. 
	A.F.V gratefully acknowledges the Institute of Theoretical Physics at the University of Graz for the warm hospitality during his visit in which part of this work was developed. A.F.V is also grateful to CP3-Origins at the University of Southern Denmark for extended hospitality during the final stages of this work.

	\appendix
	
	\section{Heat-kernel trace of the squared Dirac operator in spaces with torsion
		and vanishing curvature \label{app:heat_kernel_comput}}

	\noindent Our goal is to compute the following heat-kernel trace:
	\begin{align}
		K_\tau = \Tr \left[ \exp(-\tau \,(-\slashDNew^2) )\right] \,.
	\end{align}
	Here, we focus on the approximation of vanishing curvature and homogeneous axial-torsion. Thus, with the differential operator $-\slashDNew^2$ defined in Eq. \eqref{eq:DNew_Rzero}.
	
	Within the current setting, we can write a Fourier space representation for the heat-kernel trace:
	\begin{align}
		K_\tau = v_4 \,e^{- \tau A^2/64}  \int_q e^{-\tau \, q^2} 
		\,\textmd{tr} \left[  \exp\left(  -\frac{\tau}{4}\gamma_5 \, \sigma^{\mu\nu}A_\mu \,q_\nu \right)  \right] \,,
	\end{align}
	where $\textmd{tr}$ stands for trace over the Dirac space only. 
	\noindent To compute the momentum space integrals, we consider an expansion of the exponential inside the remaining trace:
	\begin{align}
		\textmd{tr} \left[  \exp\left(  -\frac{\tau}{4}\gamma_5 \, \sigma^{\mu\nu}A_\mu \,q_\nu \right)  \right] = 4 +  \sum_{n=1}^{\infty}\frac{(-1)^n\tau^n}{2^{2n} \, n!} \, \textmd{tr} \left[  \gamma_5 \, \sigma^{\mu_1\nu_1}  \cdots \gamma_5 \, \sigma^{\mu_n\nu_n} \right] A_{\mu_1} \cdots A_{\mu_n} \, q_{\nu_1} \cdots q_{\nu_n} \,,
	\end{align}
	where we used $\textmd{tr}\textbf{1} = 4$ in the zeroth order term.
	It turns out that only contributions with even values of $n$ are nonzero. For these terms, the trace will have an even power of $\gamma_5$, which can be combined into an identity matrix. After relabeling $n\mapsto 2n$, we find
	\begin{align}\label{eq:trace1}
		\textmd{tr} \left[  \exp\left(  -\frac{\tau}{4}\gamma_5 \, \sigma^{\mu\nu}A_\mu \,q_\nu \right)  \right] = 4 +  \sum_{n=1}^{\infty}\frac{\tau^{2n}}{2^{4n} \, (2n)!} \, \textmd{tr} \left[ \sigma^{\mu_1\nu_1}  \cdots \sigma^{\mu_{2n}\nu_{2n}} \right] A_{\mu_1} \cdots A_{\mu_{2n}} \, q_{\nu_1} \cdots q_{\nu_{2n}} \,.
	\end{align}
	Plugging \eqref{eq:trace1} back into the original integral, we find
	\begin{align}
		&\int_q e^{-\tau \, q^2} 
		\,\textmd{tr} \left[  \exp\left(  -\frac{\tau}{4}\gamma_5 \, \sigma^{\mu\nu}A_\mu \,q_\nu \right)  \right] 
		= 4 \int_q e^{-\tau \, q^2} + \sum_{n=1}^{\infty}\frac{\tau^{2n}}{2^{4n} \, (2n)!} \, \textmd{tr} \left[ \sigma^{\mu_1\nu_1}  \cdots \sigma^{\mu_{2n}\nu_{2n}} \right] A_{\mu_1} \cdots A_{\mu_{2n}} \,\int_q q_{\nu_1} \cdots q_{\nu_{2n}}\, e^{-\tau \, q^2}  \,.
	\end{align}
	The result for $\textmd{tr} \left[ \sigma^{\mu_1\nu_1}  \cdots \sigma^{\mu_{2n}\nu_{2n}} \right]$ will be a linear combination of products of the flat metric $\delta_{\mu\nu}$. As a consequence, for each value of $n$, the combination
	\begin{align}
		\mathcal{I} =\textmd{tr} \left[ \sigma^{\mu_1\nu_1}  \cdots \sigma^{\mu_{2n}\nu_{2n}} \right] A_{\mu_1} \cdots A_{\mu_{2n}} \,\int_q q_{\nu_1} \cdots q_{\nu_{2n}}\, e^{-\tau \, q^2} ,
	\end{align}
	can be rearranged into the form
	\begin{align}
		\mathcal{I} = \sum_{m_1=0}^{n} \sum_{m_2=0}^{2n} c_{m_1,m_2} \,\delta_{n,m_1 + m_2}\, (A^{2})^{n-m_1} 
		\int_q (q \cdot A)^{2m_1} \, (q^2)^{m_2} \, e^{-\tau \, q^2}\,,
	\end{align}
	for a given set of coefficients $c_{m_1,m_2}$.
	By standard covariance arguments, we can rewrite the remaining integrals as (see, \eg, App. A from \cite{Laporte:2022ziz})
	\begin{align}
		\int_q (q \cdot A)^{2m_1} \, (q^2)^{m_2} \, e^{-\tau \, q^2} = 
		\frac{\Gamma\!\left(m_1+\frac{1}{2}\right)}{\sqrt{\pi} \, \Gamma(m_1+2)}  (A^{2})^{m_1} \, \int_q  (q^2)^{m_1+m_2} \, e^{-\tau \, q^2} \,.
	\end{align} 
	It then follows that
	\begin{align}
		\mathcal{I} =  \mathcal{C}_n \, (A^{2})^{n} \, \int_q  (q^2)^n \, e^{-\tau \, q^2}\, ,
	\end{align}
	where
	\begin{align}\label{eq:Cn}
		\mathcal{C}_n = \sum_{m_1=0}^{n} \sum_{m_2=0}^{2n} c_{m_1,m_2} \,\delta_{n,m_1 + m_2}\, 
		\frac{\Gamma\!\left(m_1+\frac{1}{2}\right)}{\sqrt{\pi} \, \Gamma(m_1+2)} \,, \quad (n>0).
	\end{align}
	Notably, this expression can be computed order by order with \textit{Mathematica}. Using the function \textit{FindSequenceFunction}, we arrive at the following expression for $\mathcal{C}_n$,
	\begin{align}
		\mathcal{C}_n = \frac{8}{\sqrt{\pi }} 
		\frac{\, \Gamma \left(n+\frac{3}{2}\right)}{ \Gamma (n+2)} \,.
	\end{align}
	Going back to the original integral, we find
	\begin{align}
		\int_q e^{-\tau \, q^2} 
		\,\textmd{tr} \left[  \exp\left(  -\frac{\tau}{4}\gamma_5 \, \sigma^{\mu\nu}A_\mu \,q_\nu \right)  \right] 
		=  \sum_{n=0}^{\infty}\frac{\mathcal{C}_n}{2^{4n} \, (2n)!} \, 
		\tau^{2n}\, (A^{2})^{n} \, \int_q  (q^2)^n \, e^{-\tau \, q^2} \,,
	\end{align}
	where we have defined $\mathcal{C}_0=4$ (which is compatible with \eqref{eq:Cn} in the limit $n\to0$) to group all terms into a single sum. Finally, we can compute the remaining integral over $q$:
	\begin{align}
		\int_q  (q^2)^n \, e^{-\tau \, q^2} = \frac{1}{16 \pi^2}  \int_0^{\infty} dq^2  \, (q^2)^{n+1} \, e^{-\tau \, q^2} = \frac{\Gamma(n+2)}{16 \pi^2\,\tau^{n+2}} \,.
	\end{align}
	Therefore, we find
	\begin{align}
		\int_q e^{-\tau \, q^2} 
		\,\textmd{tr} \left[  \exp\left(  -\frac{\tau}{4}\gamma_5 \, \sigma^{\mu\nu}A_\mu \,q_\nu \right)  \right] 
		=  \frac{1}{16\pi^2 \, \tau^2} \, \sum_{n=0}^{\infty}\frac{\mathcal{C}_n\, \Gamma(n+2)}{2^{4n} \, (2n)!} \, 
		(\tau \,A^{2})^{n}  \,.
	\end{align}
	Based on \eqref{eq:Cn}, we can use \textit{Mathematica} to show that the sum in the last expression converges to
	\begin{align}
		\sum_{n=0}^{\infty}\frac{\mathcal{C}_n\, \Gamma(n+2)}{2^{4n} \, (2n)!} \, 
		(\tau \,A^{2})^{n} = \frac{1}{8} e^{\tau \,A^{2}/64} (32 + \tau \,A^{2}).
	\end{align}
	Going back to the heat-kernel trace, we finally find
	\begin{align}
		\,K_\tau = \frac{v_4}{16\pi^2 \, \tau^2}\,  \left(4 + \frac{1}{8}\,\tau \,A^{2} \right) \,.
	\end{align}
	
	
	\newpage
	\bibliography{ChSBRC}
\end{document}